# Implementation of Artifact Detection in Critical Care: A Methodological Review

Shermeen Nizami [1], Student Member IEEE; James R. Green [1], SMIEEE; Carolyn McGregor [2], SMIEEE

[1] Department of Systems and Computer Engineering, Carleton University, Canada
[2] Faculties of Business and IT, and Health Sciences, University of Oshawa Institute of Technology, Canada

*Abstract* - **Artifact Detection (AD) techniques minimize the impact of artifacts on physiologic data acquired in Critical Care Units (CCU) by assessing quality of data prior to Clinical Event Detection (CED) and Parameter Derivation (PD). This methodological review introduces unique taxonomies to synthesize over 80 AD algorithms based on these six themes: (1) CCU; (2) Physiologic Data Source; (3) Harvested data; (4) Data Analysis; (5) Clinical Evaluation; and (6) Clinical Implementation. Review results show that most published algorithms: (a) are designed for one specific type of CCU; (b) are validated on data harvested only from one OEM monitor; (c) generate Signal Quality Indicators (SQI) that are not yet formalised for useful integration in clinical workflows; (d) operate either in standalone mode or coupled with CED or PD applications (e) are rarely evaluated in real-time; and (f) are not implemented in clinical practice. In conclusion, it is recommended that AD algorithms conform to generic input and output interfaces with commonly defined data: (1) type; (2) frequency; (3) length; and (4) SQIs. This shall promote (a) reusability of algorithms across different CCU domains; (b) evaluation on different OEM monitor data; (c) fair comparison through formalised SQIs; (d) meaningful integration with other AD, CED and PD algorithms; and (e) real-time implementation in clinical workflows.**

## I. INTRODUCTION

Physiologic signals exhibit trends, dynamics and correlations reflecting the complexity of underlying patient physiology [1, 2]. Continuous monitoring of physiologic data assists clinicians in making diagnoses and prognoses in Critical Care Units (CCU), including Intensive Care (ICU), Paediatric Intensive Care (PICU), Neonatal Intensive Care Units (NICU), and the Operating Room (OR) [3]. Clinical Event Detection (CED) techniques analyze these data to identify clinically significant events and early onset indicators of various pathophysiologies as in [4-16]. Parameter Derivation (PD) techniques derive clinically useful low frequency parameters from high frequency input data as in [17-23].

Artifacts are extraneous signals with randomly varying amplitudes, frequencies and duration that interfere with physiologic signals acquired in clinical settings [24]. Longitudinal studies [25-28] infer that Original Equipment Manufacturer (OEM) patient monitors have relatively simplistic built-in data preprocessing for Artifact Detection (AD). In this paper, the term AD encompasses one or more of the following mechanisms: (a) identification, detection or annotation of artifacts, (b) elimination of artifacts or artifact-laden data and (c) suppression or filtering of artifacts e.g., by using ensemble averaging or adaptive filtering [29]. The terms *noise* and *cleaning* have been used in publications as alternatives for the terms artifacts and AD respectively. OEM monitors typically come with a black box approach to preprocessing artifacts in physiologic data. Clinicians debate the reliability of OEM monitor data as they deduce that some built-in algorithms employ a reductionist approach that oversimplifies complex human physiology [30]. Artifacts can mimic physiologic data [31-33], adding to the challenge of distinguishing between the two. As a result, data logged by OEM monitors remains impacted by artifacts [34-36]. This increases false alarm rates in monitors [36-38], which leads to staff desensitization [30, 39, 40]. Clinicians cannot rely on analysing artifact-laden monitor data [41], which in the past has resulted in incorrect diagnoses [33, 42]; unnecessary therapy; surgery and iatrogenic diseases [32]. Independent research groups have developed a variety of post processing techniques to address the problem of artifacts in OEM monitor data. Fig. 1 shows the AD configurations reviewed in this paper. The purpose of AD is identical in any configuration, i.e., to assess and enhance the quality of physiologic data. Fig. 1 depicts Signal Quality Indicators (SQI) among other output variables. Latest AD research shows increasing interest in devising SQIs [43-46]. This paper conducts a critical review of the development and utility of SQIs.

This methodological review introduces six thematic taxonomies for a critical analysis of the state-of-the-art in AD. The objective is to synthesize the status of clinical evaluation and implementation of AD in various CCUs. Reviewed publications are listed in Table I. Section II discusses past reviews of AD. Section III briefly describes research methodology followed by detailed development of a unique taxonomy for each of the six themes. Review results are thematically synthesized in Table II. Section IV concludes the review by highlighting open research problems. It also provides specific recommendations for new research directions in promoting implementation of AD in real-time clinical workflows.

## II. RELATED WORK

Historically, AD has been reviewed with significantly different scope and context from this paper, which makes this review a novel contribution in this research space. Borowski *et al.* [47] have conducted a comprehensive review on alarm generation in medical devices. Their primary focus is ergonomics and human factors engineering, although, they briefly synthesize some AD algorithms and, as a result, recommend using multivariate alarm systems. Schettlinger *et al.* [48] have largely reviewed their own research in developing univariate filters for outlier, trend and level shift detection in various ICU data types. They extensively describe filter development and compare advantages and disadvantages of different filter designs. Chambrin [49] infers that multivariate techniques can reduce false alarm rates in ICUs. Numerous publications, such as [37, 43, 48, 50-52], provide comparative numeric summaries of performance metrics of AD algorithms. This review neither reiterates algorithmic details nor performance comparison as that falls outside its scope. Takla *et al.* [28] note that while AD techniques developed by independent researchers may have higher specificity than built-in algorithms in OEM monitors, extensive studies are required to evaluate their accuracy prior to real-time implementation in clinical environments. Siebig *et al.* [38] demonstrate agreement amongst clinical staff, including intensivists, that integrative monitoring through data fusion can potentially yield better results as compared to simpler univariate threshold detection methods. Imhoff *et al.* [37, 52] emphasize that methodological research is needed for integrating multivariate AD algorithms in real-time clinical systems. However, they do not present a conceptual framework to conduct such research.

## III. METHODOLOGICAL REVIEW

Papers of interest were located in Scopus, IEEE Xplore and PubMed databases using the keywords artifact, artefact, artifact detection, alarm, physiologic monitoring, intensive care, neonatal intensive care, operating room and patient monitors amongst others. No constraints were applied on the publication year. Google Scholar was also used. Literature was then methodologically reviewed using six thematic taxonomies developed in this section: (1) Critical Care Unit, (2) Physiologic Data Source, (3) Harvested Data, (4) Data Analysis, (5) Clinical Evaluation and (6) Clinical Implementation. Fig. 2 depicts an overview of the thematic taxonomies.

### A. Critical Care Unit

This theme catalogues domain specific AD research to examine its reusability and portability across other CCUs. The taxonomy developed for this theme is: ICU, PICU, NICU, OR and Other relevant studies. This taxonomy, tabulated in Table I, reveals that almost all techniques are evaluated in a single domain, with the exception of [23] which is evaluated in the ICU, PICU and OR. In general, trends in physiologic signals display similar dynamics across CCU domains [53]. This drives the hypothesis that AD algorithms could be modified for use across different critical care settings. However, domain specific information such as types and frequency of data, and patient demographics such as range of age, weight and medical condition may be directly, or indirectly, hard coded in the algorithm. Therefore, it is important to consider the original domain of clinical application prior to attempting application elsewhere.

### B. Physiologic Data Source

This theme synthesizes the methodologies used to source physiologic data. The theme taxonomy is: Physiologic Monitor, Inclusion/Exclusion Criteria and Sample Size, as tabulated in the second column of Table II. The types, frequency, numeric values and quality of data produced by patient monitors (and probes) differs between models from the same or different OEMs. This is due to different built-in proprietary signal preprocessing, inclusive of AD [28, 54]. For example, comparative studies show characteristic discrepancies in neonatal data acquired using various OEM Pulse Oximeters (PO) including Masimo SET Radical (Masimo Corp., Irvine, CA, USA), Datex-Ohmeda TruSat (GE Healthcare, Chalfont St Giles, UK), Siemens SC7000 (Siemens UK, Frimley, UK), Nonin 7500 (Nonin Medical, Plymouth, MN, USA), [55]; Nellcor OxiMax N-600x (Covidien-Nellcor, Boulder, CO, USA) [55, 56]; Philips FAST MP50 [56] and Philips Intellivue MP70 (Philips, Germany) [57]. Therefore, knowledge of the monitor OEM and model is important for recognizing different biases introduced in the data. Some studies may collect data under certain inclusion and exclusion criteria as shown in Table II. Sample size is almost always given in the literature. The discretion generally lies with the researcher to determine if enough data is available to draw a statistically and clinically sound decision. Sample size may be deduced based on trial, availability of resources and study requirements.

### C. Harvested Data

The taxonomy developed under this theme is: Data Type and Acquisition/Sampling or Storage Frequency. In Table II, the top half of each cell in column three summarizes the types and frequencies of data harvested in each study. Knowledge of the OEM is also a convenient indicator of this information. Routinely harvested or monitored physiologic data types in critical care include: Electrocardiograph (ECG), Heart Rate (HR), Breathing or Respiratory Rate (RR), Impedance Respiratory Wave (IRW), Noninvasive Oxygen Saturation ($SpO_2$), Invasive or noninvasive Arterial Oxygen Saturation ($SaO_2$), Temperature (Temp), the set of Blood Pressure (BP) measurements (namely Systemic Artery, Pulmonary Artery, Central Venous, Systolic (SBP), Diastolic (DBP) And Mean (MBP) Pressures, Arterial (ABP)), Maximal Airway Pressure (MAP), Expired Air Volume (EV), Minute Ventilation (MV) and Transcutaneous Partial Pressures of Carbon Dioxide ($TpCO_2$) and Oxygen ($TpO_2$). Frequency of the data is restricted by analog to digital sampling capability of the monitor [58]. Storage frequency depends on data logging capabilities of both the OEM monitor and the hardware and software mechanism used for storing study data. The synthesis in Table II demonstrates that physiologic data are acquired, derived, sampled and stored at varying frequencies. It is common for AD algorithms to be hard coded to input particular types of data streams having specific frequencies. High frequency signals, such as continuous waveforms of ECG, ABP and PPG, are sampled at 100 Hz or more. Data that form low frequency time series, such as HR, SBP, DBP and $SpO_2$, are either time-averaged at a rate of once every second to once every minute from high frequency signals; or measured intermittently every half hour to an hour such as temperature and non-invasive BP.

### D. Data Analysis

Analytic aspects of AD algorithms are reviewed using the following thematic taxonomy: Dimensionality, Focus, Signal Quality and Clinical Contribution. Theme findings are summarised in Table II, in the bottom half of each cell in column three. Dimensionality represents the number of variables or data types that an algorithm is capable of analyzing. According to Imhoff et al. [37], both the clinical problem and the approach to solving it are (a) univariate when a single feature of a specific data stream is analysed; or (b) multivariate when results are derived from simultaneous analysis of multiple variables; and in between these two approaches lays (c) the univariate clinical problem solved using multivariate data. A new definition for multivariate analysis has emerged in recent research, and as such should be appended to the above as: (d) an algorithm is multivariate if it analyses different metrics derived from the same physiological signal. Examples of multivariate type (d) research on the PPG is found in [59] and on multi-lead ECG in [34, 50, 60-63] A uniquely different approach to multivariate analysis is found in [64, 65], where two different data streams were acquired from the same probe making their physiologic correlation easier to exploit.

The Data Analysis theme characterizes the focus of each algorithm as: (1) stream; (2) patient; and/or (3) disease-centric. A stream-centric algorithm aims to indicate, quantify or improve signal quality of the data stream for increased reliability. Although the term patient-centric has broad implications in health care, it means here that the algorithm was trained on patient specific data and is therefore heavily tailored to each sample patient in the study. For example, baseline data from a particular patient may be required to instantiate an algorithm. BioSign [66] is an example of a real-time, automated, stream and patient-centric system. It produces a single-parameter representation of patient status by fusing five dimensions of vital sign data. A disease-centric methodology focuses on identifying or predicting a specific disease or a clinically significant outcome. The bradycardia detector in [67] and the prognostic tool for Late Onset Neonatal Sepsis (LONS) in [68] are both examples of stream and disease-centric approaches. The clinical contribution taxonomy reveals two unique configurations of AD: (i) Standalone AD and (ii) Coupled AD. As Fig. 1 illustrates, standalone AD techniques typically output filtered or original physiologic data, annotations and SQI. Standalone techniques are labeled as AD under clinical contribution in Table II. Coupled AD techniques are coded as part of algorithms that identify and/or filter artifacts similar to standalone AD techniques with the additional ability for CED or PD. Coupled AD configuration is shown in Fig. 1 and documented under clinical contribution in Table II.

Signal Quality is a key element in this taxonomy. Missing segments, error, noise and artifacts inevitably affect data quality thus adversely impacting analytic accuracy and reliability [69]. To address this issue, Clifford et al. [70] recommend that an SQI calibrated to provide a known error rate for a given value of the SQI be made available for each datum. Nizami et al. [71] infer that it suffices for SQIs to be available at a frequency relative to the requirement of another AD, CED or PD application that consume the SQIs. The latter is particularly relevant when down sampled data is required by CED or PDs.

It is hypothesized that the performance of post processing AD algorithms can improve by consuming streaming SQIs output in parallel for each data stream by patient monitors [72]. However, use and delivery of SQIs is not yet standardised across OEM monitors. For example, in case of a lead disconnection the Philips IntelliVue (Philips, Germany) monitor outputs a random value above 8 million in data streams such as the ECG, IRW, HR, RR, SpO$_2$ and BP. This also generates a corresponding alert type of 'medium priority technical alarm' logged at a value of "2". However, this value is qualitative and not quantitative. The Delta, Gamma, Vista, Kappa and SC6002 - SC9000 monitor series (Dräger Medical Systems, Lubeck, Germany) output an SQI value between 0-100% for the electroencephalography (EEG) channel. This SQI is calculated using sensor impedance data, artifact information and other undisclosed variables. The same monitors also output a fixed label 'ARTF' indicating artifact on any monitored data stream [73, 74]. For example, the monitor classifies QRS complexes only at ECG values > 0.20 mV for widths > 70ms. An artifact condition 'ARTF' may be declared when the ECG signal does not meet these minimum criteria. Although OEM monitors may output signal quality information, there is no logical way to compare the SQIs produced by different monitors. Two confounding reasons are: (1) difference in quantification of SQI; and (2) lack of literature on proprietary algorithms. This reduces researchers' ability to determine how data has been affected from acquisition to logging. As a result, a post processing AD algorithm may need to be strictly matched to input data sourced from a particular OEM monitor as shown in Fig. 1. Comparative studies led by Masimo [75, 76] declare that its 510 (k) FDA approved RADICAL SET technology has the highest quality measure called *Performance Index* (*PI*) as compared to 19 other OEM POs. In these publications, Masimo defines and calculates *PI* as the percentage of time during which a PO displayed a current SpO$_2$ value that was within 7% of the simultaneous control value. However, the Masimo SET technology does not automatically evaluate or log this quantitative SQI. Another quality measure called *Dropout rate* (*DR*) was calculated in [76], which equals the percentage of measurement time during which no current SpO$_2$ values are displayed. Although Masimo SET showed equal or worse *DR* than two Datex-Ohmeda POs, the reasons for this data loss are not discussed by Masimo. Independent research groups that compared OEM POs in [55-57] neither researched the effect of the difference in data characteristics on SQIs nor did they mention if the POs output SQIs. Through a recent discourse in [77] on historic developments of the Masimo SET technology, its OEM has replied to Van Der Eijk et al. [56], claiming greater accuracy in unstable conditions, such as motion artifact and low perfusion, leading to lower false alarm rates. However, [77] does not describe any SQIs that can be consumed meaningfully by other AD, CED or PD applications. This review recommends that AD algorithms that produce SQIs, such as *PI* and *DR* amongst others, be evaluated upon data acquired in [55-57] to contribute towards future AD research.

It follows that AD algorithms designed to post process OEM monitor data must also consume and deliver standardised SQIs. In this way another AD algorithm or a CED or PD mechanism can make informed choices concerning data quality and validity. Review results in Table II show the increasing trend in SQI development. However, no framework exists to uniformly deliver, compare or combine these SQIs for integration with clinical workflows.

### E. Clinical Evaluation

Patient safety requires clinical evaluation of algorithms prior to real-time clinical implementation. This theme reviews clinical evaluation methods bases on this taxonomy: Data Annotation, Mode and Performance Metric. Results are given in Table II, in the top half of each cell of column four. There are no rules that define gold standards for clinical evaluation of AD performance. Each study sets its own gold standard against which its performance is evaluated. This includes evaluations of OEM monitors. Annotated physiologic data, where available, typically serve as the gold standard for validation studies. Events of interest in the data, such as artifacts and clinically significant events, are marked in real-time or retrospectively. The onus of perceiving what constitutes an "event" is on the expert reviewers, who identify the event to the best of their knowledge. Inter-reviewer variability is the significant [78] or subtle [27, 79] difference known to arise when the same dataset is annotated by different reviewers. Retrospective data annotation has been supported by video monitoring in [27, 31, 36]. Video monitoring is only useful when the event is visually perceptible such as sleep movements, certain seizures and routine care. However, it cannot capture crucial physiologic changes such as HR deceleration or BP elevation. The advantage of real-time annotations is recording of richer and more accurate content with input from staff on duty. However, this can be costly and requires cooperation from busy staff. Study data is either collected in real-time from patient monitors or acquired in an "offline" mode for secondary analysis from existing databases. Review results show that majority of AD techniques were validated on offline patient data and very few were tested in real-time CCU environments. Table II also documents the types of performance metrics used in each research. It shows the common trends that will help future researchers to design and compare different algorithms by evaluating them using the same metrics. Numeric comparison of these performance metrics can be found elsewhere in [37, 43, 48, 50-52]. Performance metrics need to be interpreted very carefully since statistical significance, or absence thereof, is not always representative of clinical significance, or absence thereof. For example, one missed clinical event may not signify a statistical difference in the sensitivity of one OEM monitor over another. However, the same event could be very important clinically and crucial that it not be missed even once. Theoretically, a missed event or a false alarm are caused by artifacts of various types; for example, motion artifact and power line or optical noise induced in an attached or detached sensor. Studies that collected real-time annotated data or video monitored data, such as [27, 31, 36, 80-86] among others, can utilise the same data sets to develop and validate SQIs. Compatible SQIs can be used to compare performance of different AD algorithms and OEM monitors. Performance metrics, for example sensitivity and specificity in alarm studies, can be re-evaluated taking into consideration the SQI at each alarm instance

### F. Clinical Implementation

This theme reviews the clinical implementation status of AD techniques. Theme results are given in Table II, in the bottom half of each cell in column four. This review reveals that the vast majority of AD techniques that are published have not been put into clinical practice. This section critically reviews implementation of some commercialised OEM monitors and the very few techniques developed by independent research groups that made their way into clinical workflows.

The Philips IntelliVue monitoring system (Royal Philips Electronic, Netherlands) features *Guardian Early Warning Score* (EWS) allowing each hospital to choose its own scoring criteria; *Neonatal Event Review* which detects apnoea, bradycardia and desaturation; *Oxy-cardiorespirography* (Oxy-CRG) with compressed trends of a neonate's HR, RR, and $SpO_2$; and *ProtocolWatch* that is claimed to reduce sepsis mortality rates. Presumably Intellivue preprocesses patient data for artifacts prior to CED or PD, however, no validation studies or algorithm details of this system are published. GE Intellirate[TM] monitor (Milwaukee, WI, USA) issues asystole, bradycardia and tachycardia alerts by fusing ECG, ABP and PO data. It was evaluated by GE on a small population of 55 CCU patients in 2002 [87]. The evaluation was critiqued in [88] for lack of description of patient demographics and algorithm specifications. GE has republished the exact same study in 2010. The Saphire clinical decision support system [89] uses Intellirate[TM] technology, but does not evaluate it. Multi-lead ECG arrhythmia detection is deployed by GE in Datex-Ohmeda Bedside Arrhythmia Monitoring (Milwaukee, WI, USA), MARS Ambulatory ECG system and MARS Enterprise (Freiburg, Germany). MARS uses the OEM's EK-Pro Arrhythmia Detection Algorithm which has been evaluated in over 2000 monitored hours spanning at least 100 patients. Surely, these techniques fall under the category of AD coupled with CED and PD. However, literature lacks comparison between different models marketed by the same or different OEMs.

OEM Covidien-Nellcor (Boulder,CO, USA) has developed $RR_{oxi}$, a coupled AD and PD technology that derives RR from PO. $RR_{oxi}$ has been validated in real-time on 139 healthy subjects in [18]. The OEM is commended for this substantial evaluation. However, healthy subjects are not representative of patient populations which the device is intended to monitor. $RR_{oxi}$ has been validated retrospectively in 12 patients with congestive heart failure, by evaluating 20 minutes of data from each patient [19]. However, larger studies that investigate patient populations with several different pathophysiologies are required to convince clinicians to adopt another patient monitoring technology in their workflows. Fidelity 100 is an FDA 510 (k) approved wireless ECG monitor developed by Signalife (Studio City, CA, USA), which was evaluated in real-time in 54 patients undergoing percutaneous coronary intervention [83]. Although these monitors come with different settings applicable for use in different types of CCUs, validation studies on population data from all application domains are not found. There are a growing number of online open source physiologic databases, such as Capnobase [90, 91], FDA ECG Warehouse [92], hemodynamic parameter database [93], and PhysioNet [94]. It is recommended that these databases be used to compare and validate OEM monitors of different makes and models.

Rest of this section reviews clinical implementation of AD research developed by independent research groups. CIMVA (Therapeutic Monitoring Systems Inc., Ottawa, Canada) is a patented multi-organ variability analytics technology developed by Seely *et al.* [95-98]. It is an online tool comprising of multiple coupled AD and CED algorithms with SQIs. Its AD performance is evaluated in [97]. CIMVA research can benefit from the recommendations made in the next section regarding common interfaces and formalised SQIs. This will allow for new AD, CED and PD algorithms to be integrated and tested as part of the CIMVA architecture. Otero *et al.* have implemented TRACE, a graphical tool which allows clinicians to edit monitoring rules and criteria in real-time. Coupled AD and CED algorithms based on fuzzy set theory input these customized criteria to generate patient alarms in [81, 99], and detect sleep apnoea in [100]. Given its promising results, evaluation of TRACE against similar OEM monitors is recommended. The research conducted in 1999 by Schoenberg *et al.* [54] was integrated as part of the commercially available iMDsoft Clinical Information System [101]. However, algorithmic details and evaluation were never published. Artemis is a real-time data analytics system currently undergoing clinical evaluation in multiple NICUs around the globe [102-104]. As part of the Artemis framework, Nizami *et al.* [71] present SQI processing to improve the performance of coupled AD and CED algorithms for LONS. Ongoing Artemis research includes AD [80]; CED of Apnoea of Prematurity [5, 15, 105]; as well as pain management [9]. Adoption of the structured approach to AD recommended in this review can enhance the clinical performance of coupled AD and CED in Artemis. The coupled AD and CED algorithm for Bradycardia by Portet *et al.* [67] was evaluated on offline NICU data with the intent of integration with the BabyTalk project. BabyTalk's proof of concept has been described in several publications [106-110]. However, latest research by Hunter *et al.* in 2012 [111] infers that a long road lies ahead, including necessary clinical trials, before BabyTalk could be implemented in real-time clinical workflows. Several new AD and CED algorithms can be tested to improve outcomes of this project by incorporating formalised interfaces as recommended in this review. The patented HeRO™ system (Medical Predictive Science Corp.,

Charlottesville, VA, USA) that scores neonatal Heart Rate Variability (HRV) for predicting LONS is developed using coupled AD and CED algorithms [68]. It conducts a multivariate type (d) analysis of the ECG with multiple coefficients yielding a more sensitive result [84]. The algorithms in this 510 (k) FDA approved device have been extensively described and evaluated both offline and in real-time by Moorman *et al.* [84, 112-122]. This pioneering research has shown promising reduction in neonatal mortality by 2% in a randomized control trial on 3003 preterm babies across nine NICUs in the US [114]. The drawback of this trial was a 10% increase in blood work and 5% more days on antibiotics in the monitored infants. Ironically, this constitutes the original problem this research set out to resolve in [119]. It is recommended that other variability measures, such as those of RR as in [10] and PPG as in [123], be evaluated in comparison with the HeRO™ HRV score to come up with (a) individual scores for each data type; and (b) a composite score that exploits sensor fusion for improved outcomes. BioSign [66] has been evaluated retrospectively and in real-time in a number of clinical studies including randomized control trials in Europe and the US before 2006. However, no later publications could be located.

## IV. CONCLUSIONS

This section derives conclusions from the thematic review. Post processing AD techniques are highly domain specific. This necessitates modification for validation and reuse in a different CCU domain. Algorithms may be hard coded to input OEM specific data types and frequency. This limits their use with different OEM monitor data. They may be validated under certain inclusion/exclusion criteria which need to be considered when applying the techniques in other contexts. Acquisition and sampling frequency play an important role in patient management since a critically ill patient's condition may deteriorate to a life threatening extent within seconds. Reusability is deterred when such implicit limitations are not expressed. Therefore, adoption of a standardised structured approach for design and reporting of standalone and coupled AD research is recommended. Conformity to generic input and output interfaces will ensure presence of all pertinent information. These interfaces, with common definitions for data type, frequency, length and SQIs, shall allow for matched selection and composition with other AD, CED or PD algorithms. Results from the first three themes are useful in selecting one or more AD algorithms that fulfill data requirements of given CED or PD techniques. Selected AD algorithms can be mixed and matched to discover optimal compositions for varying clinical requirements.

OEM monitors marketed for use across different CCUs have undisclosed built-in preprocessing algorithms, inclusive of AD. Moreover, studies validating their use in different CCU domains and patient population are scarce. The resulting unknown bias imparted in OEM data leads to inevitable variance in analytic results which can effect clinical decisions. This variability can be decreased if monitors output comparable standardised SQIs. As of yet, SQIs do not conform to any standards and are derived differently in each publication, whether it be SQIs delivered by OEM patient monitors or by AD algorithms developed by independent research groups. Interestingly, none of the reviewed algorithms reported using SQIs provided by OEM monitors. Clinical utility of SQIs can be enhanced by using formalised definitions such that SQIs output by different preprocessing and post processing AD algorithms are comparable as well as compatible. An SQI matched to the same set of definitions is also proposed as a requirement at the input of AD, CED or PD algorithms. The objective is to enable the AD, CED or PD algorithm to compare the incoming SQIs generated by one or more AD techniques with their required SQI value. The CED or PD algorithm may then accept or reject incoming data segments based on fulfillment of the required SQI value. In conclusion, standardised SQIs are vital to allow informed clinical choices concerning use and validity of physiologic data.

Results of the Clinical Evaluation theme show that majority of AD techniques are validated on offline data and very few have been evaluated in real-time CCU environments. Clinical Implementation theme reveals that AD techniques developed by independent research groups have rarely found their way into clinical implementation. This leads to the inference that a gap exists between research efforts in AD and their utilization in real-time clinical workflows.

Whereas real-time clinical implementation of AD algorithms is noticeably lacking, there is growing interest amongst clinicians to use CED and PD for automated clinical decision support in CCUs, such as in [8, 104, 124-135]. Physiologic signal quality assessment through integration of AD can improve the outcome, reliability and accuracy of CED and PD research. The conclusions and recommendations of this review provide new research direction for promoting integration of AD in real-time clinical workflows.

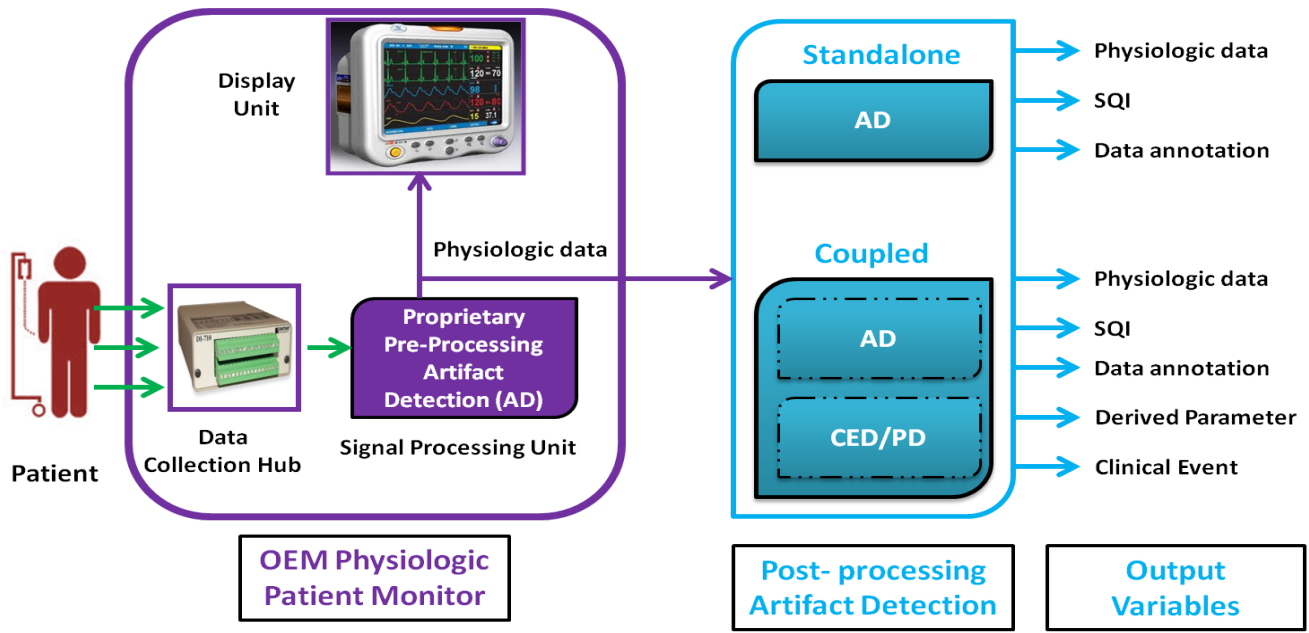

Fig. 1. Patient data acquisition by an OEM physiologic monitor with built-in preprocessing inclusive of AD, followed by post-processing AD, either standalone or coupled with CED and/or PD, with output variables resulting from the entire data analysis.

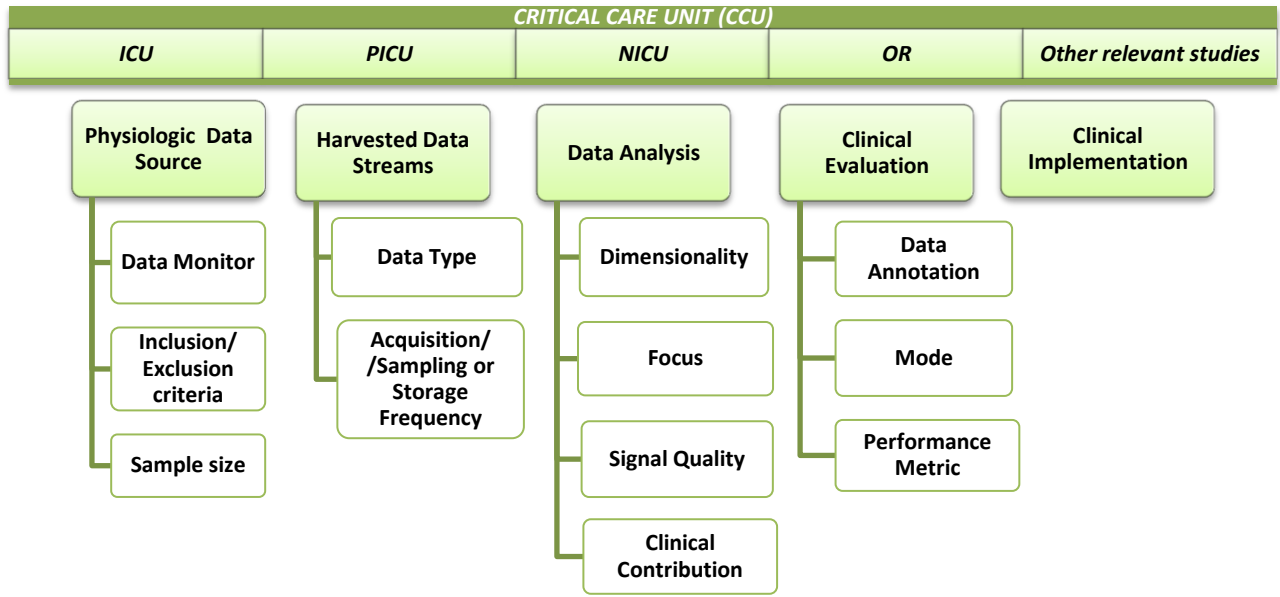

Fig. 2.  Overview of the Six Thematic Taxonomies

TABLE I: CRITICAL CARE UNIT (CCU)

| Theme I: Critical Care Unit | | | | |
|---|---|---|---|---|
| *ICU* | *PICU* | *NICU* | *OR* | *Other relevant studies* |
| **Bradley**, 2012 [97]<br>**Hu**, 2012 [136]<br>**Li**, 2012, 2009, 2008 [20, 21, 59]<br>**Sun**, 2012 [137]<br>**Scalzo**, 2012 [60]<br>**Borowski**, 2011 [138]<br>**Siebig**, 2010 [27]<br>**Schettlinger**, 2010 [51]<br>**Charbonnier**, 2010, 2004 [139, 140]<br>**Otero**, 2009 [81]<br>**Blum**, 2009 [141]<br>**Aboukhalil**, 2008 [88]<br>**Otero**, 2007 [99]<br>**Sieben**, 2007 [82]<br>**Zong**, 2004 [142]<br>**Jakob**, 2000 [143]<br>**Schoenberg**, 1999 [54]<br>**Ebrahim**, 1997 [22]<br>**Feldman**, 1997 [23] | **Zhang**, 2008 [144]<br>**Tsien**, 2000, 1997 [85, 145, 146]<br>**Ebrahim**, 1997 [22]<br>**Feldman**, 1997 [23] | **Monasterio**, 2012 [147]<br>**Hoshik**, 2012 [148]<br>**Nizami**, 2011 [71]<br>**Salatian**, 2011 [26]<br>**Belal**, 2011 [149]<br>**Blount**, 2010 [80]<br>**Quinn**, 2009 [150]<br>**Zhang**, 2008 [144]<br>**Portet**, 2007 [67]<br>**Walls-Esquival**, 2007 [31]<br>**Moorman**, 2006 [84]<br>**Tsien**, 2001, 2000 [145, 151, 152]<br>**McIntosh**, 2000 [64]<br>**Cao**, 1996 [65] | **Karlen**, 2012, 2011 [90, 91]<br>**Koeny**, 2012 [153]<br>**Schmid**, 2011 [36]<br>**Yang**, 2010, 2009 [53, 58, 154]<br>**Ansermino**, 2009 [155]<br>**Hoare**, 2002 [72]<br>**Gostt**, 2002 [156]<br>**Ebrahim**, 1997 [22]<br>**Feldman**, 1997 [23]<br>**Sittig**, 1990 [157]<br>**Navabi**, 1989 [158] | **Clifford**, 2012 [34]<br>**Redmond**, 2012 [44]<br>**Reisner**, 2012 [159]<br>**Martinez-Tabares**, 2012 [160]<br>**Silva**, 2012 [161]<br>**Hayn**, 2012 [61]<br>**Jekova**, 2012 [50]<br>**Di Marco**, 2012 [62]<br>**Johannesen**, 2012 [63]<br>**Addison**, 2012 [18][61]<br>**Lázaro**, 2012 [17]<br>**Otero**, 2012 [100]<br>**Bsoul**, 2011 [162]<br>**Xia**, 2011 [163]<br>**Zaunseder**, 2011 [164]<br>**Kužílek**, 2011 [165]<br>**Sukor**, 2011, [29]<br>**Acharya**, 2011 [166]<br>**Khandoker**, 2011 [167]<br>**Nizami**, 2010 [168]<br>**Nemati**, 2010 [169]<br>**Alvarez**, 2010 [170]<br>**Gil**, 2008 [171, 172]<br>**Chen**, 2008 [69]<br>**Kostic**, 2007 [83]<br>**Yu**, 2006 [173]<br>**Tarassenko**, 2006 [66] |

TABLE II: REVIEWED THEMES II – VI

| Author | Theme II: Physiologic Data Source Data Monitor/ Inclusion (I)/ Exclusion (E) criteria/ Sample size | Theme III: Harvested Data Data Type/Acquisition(A)/ Sampling or Storage (S) Frequency / Theme IV: Data Analysis Dimensionality/ Focus/ Signal Quality/ Clinical Contribution | Theme V: Clinical Evaluation Data Annotation/ Mode/ Performance Metric / Theme VI: Clinical Implementation |
|---|---|---|---|
| Bradley [97] | Phillips Intellivue MP70 (Philips Healthcare, Andover, Massachusetts)/ I: respiratory and/or cardiac failure, enrollment in the study within 36 hours of ICU admission, expected period on study greater than 72 hours/ E: chronic atrial fibrillation, transfer from another ICU/ 34 patients | ECG, EtCO$_2$ / A: ECG at 500 Hz; S: 125 Hz | Retrospective/Offline/ Percentage data loss |
|  |  | Univariate/ Stream/ Developed SQI/ AD | CIMVA (TMS Inc., Ottawa, Canada) |
| Hu [136] | Monitor not specified, data storage system BedMasterEx™/ (i) I: code blue adult patients > 18 yr (ii) I: control patients with same: (1) All Patient Refined or Medicare Diagnosis Related Group; (2) age ± 5 years; (3) gender; (4) same CCU/ (ii) E: had code blue; experienced an unplanned ICU transfer/ (i) 223 patients (ii) 1768 patients | Alarms/Not applicable | Retrospective/ Offline & Simulated real-time/ Se, FA |
|  |  | Multivariate/ Stream + disease-centric/ No SQI / AD + CED | None |
| Li [59] | [1]MIMIC II database/ I: asystole & ventricular tachycardia/ 104 patients, total 1055x6s | PPG/ A & S: 125 Hz | Retrospective/ Offline/ Acc, ROCC |
|  |  | Univariate/ Stream + Patient-centric/ Developed SQI/ AD | None |
| Sun [137] | i - [1]MIMIC II database, ii – BIOPAC PPG 1OOC module, TSD200 PPG reflective transducer, iii - BIOPAC OXY 100C module with TSD123A SpO$_2$ finger transducer / i -87 seconds, ii & iii –3 healthy volunteers x 20 sec | PPG, ECG/ A & S : 125 Hz for MIMIC II datasets;Not specified for BIOPAC data | (i ) Retrospective (ii) By real-time observer/ Offline/ RMSE |
|  |  | Multivariate/ Stream-centric/ No SQI/ AD | None |
| Scalzo [60] | I: intracranial hypertension/ 108 patients | Intracranial Pressure (ICP) & ECG waveforms, ICP alarms/ A: ICP at 240 Hz | Retrospective/ Offline/ AUROC, TPR, FPR |
|  |  | Multivariate (d)/ Stream-centric/ No SQI/ AD | None |
| Borowski [138] | Infinity patient monitoring system, Dräger Medical, (Lubeck, Germany)/ 1245:52:28 hr | SBP, MAP, HR, SpO$_2$, alarms/ S: 250 Hz | Retrospective/ Offline/ Se, FARR |
|  |  | Univariate/ Stream-centric/ No SQI/ AD | None |
| Siebig [27] | Infinity patient monitoring system & full-disclosure data logging software eData, (Dräger Medical, Lübeck, Germany)/ 38 patients, total 515 hr | At least HR, IABP, SpO$_2$, alarms/ S: 1 Hz | Retrospective with video/ Offline/ Relevant and FA |
|  |  | Univariate/ Stream-centric/ No SQI/ AD | None |
| Schettlinger [51] | 1 hr of SBP; 30 mins of HR | SBP, HR/ A & S: 1 Hz | Not specified/ Offline/ Not specified |
|  |  | Univariate/ Stream-centric/ No SQI/ AD | None |
| Charbonnier [139] | Datex-Ensgtrom monitor with a multi-parameter module/ I: various disorders with specific clinical contexts such as mechanical ventilation and cessation of sedative drug administration/ 14 patients, total 50 hr | SpO$_2$,SBP, DBP,MBP, HR, MAP, RR,EV, MV, maximal airway flow A: 100 Hz, S: 1 Hz | Retrospective/ Offline/ Se, Sp |
|  |  | Multivariate/ Stream-centric/ No SQI / AD | None |

| Author | Theme II: Physiologic Data Source Data Monitor/ Inclusion (I)/ Exclusion (E) criteria/ Sample size | Theme III: Harvested Data Data Type/Acquisition(A)/ Sampling or Storage (S) Frequency Theme IV: Data Analysis Dimensionality/ Focus/ Signal Quality/ Clinical Contribution | Theme V: Clinical Evaluation Data Annotation/ Mode/ Performance Metric Theme VI: Clinical Implementation |
|---|---|---|---|
| *Li* [20, 21] | [1]MIMIC II database/ 6000 hr | ECG, ABP/ A: 500 Hz; S: 125 Hz | Retrospective/ Offline/ True and False Alarms, RMSE |
| | | Multivariate/ Stream-centric/ Developed SQI / AD + PD | None |
| *Otero* [81] | Monitor not specified, data logging software SUTIL/ 78 patients, total 196 hr | HR, RR, BP, $SpO_2$/ Not specified | Real-time/ Real-time,TRACE [174]/ CDR, FPR |
| | | Multivariate/ Stream + disease -centric/ No SQI / AD + CED | None |
| *Blum* [141] | Solar 9000 monitors and Monitor Capture Server data logging software (GE Healthcare, UK)/ 28 days, total 293,049 alarms | SBP, MAP, CVP, Chest Impedance (CI) and their alarms/ Not applicable | Retrospective/ Offline/ Sp |
| | | Multivariate/ Stream + patient -centric/ No SQI / AD | None |
| *Aboukhalil* [88] | [1]MIMIC II database/ I: critical ECG arrhythmia alarm in the presence of one channel of ECG and an ABP waveforms/ E: 49 patients with active intra-aortic balloon pumps / 5386 alarms from 447 patients, total 41,301 hr | ECG, ABP/ A: 500 Hz; S: 125 Hz | Retrospective /Offline / True and False Alarm Reduction Rates |
| | | Multivariate/ Stream + disease -centric/ Specified SQI / AD + CED | None |
| *Otero* [99] | OEM not specified / 71 patients, total 175 hr | HR, RR, BP, $SpO_2$/ S: 1 Hz | Not specified / Offline, TRACE [174]/ CDR, FPR, FNR |
| | | Multivariate/ Stream-centric/ No SQI / AD | None |
| *Sieben* [82] | Not specified | RR, $SpO_2$, arrhythmia indicator, HR, PR, premature ventricular contraction, SBP, DBP, MBP, temperature, thresholds, alarms/ Not specified | Real-time / Offline/ Se, FARR |
| | | Multivariate/ Stream-centric/ Developed SQI but insufficient details given / AD | None |
| *Charbonnier* [140] | OEM Not specified / 18 patients, total 36 hr | HR, SBP, DBP, MBP, $SAO_2$, MAP / A: 100 Hz, $SAO_2$ at 0.2 Hz; S: 1 Hz | Real-time/ Simulated data/ Number of False, True & Technical Alarms |
| | | Univariate/ Stream-centric/ No SQI/ AD | None |
| *Zong* [142] | [3]MIMIC database / I: at least multi-lead ECG, ABP / 46 patients, total 1890 hr | ECG at 500 Hz, ABP at 125 Hz | Retrospective / Offline/ TA, FA |
| | | Multivariate/ Stream-centric/ Calculated SQI / AD | None |

| Author | Theme II: Physiologic Data Source<br>Data Monitor/ Inclusion (I)/ Exclusion (E) criteria/ Sample size | Theme III: Harvested Data<br>Data Type/Acquisition(A)/ Sampling<br>or Storage (S) Frequency<br><br>Theme IV: Data Analysis<br>Dimensionality/ Focus/ Signal Quality/ Clinical<br>Contribution | Theme V: Clinical Evaluation<br>Data Annotation/ Mode/<br>Performance Metric<br><br>Theme VI: Clinical Implementation |
|---|---|---|---|
| Jakob [143] | Datex-Ohmeda, (Helsinki, Finland)/ I: systemic and pulmonary artery catheters after coronary artery bypass grafting/ E: clinical signs of heart failure/ 41 patients | HR, systemic (SAP) & pulmonary (PAP) artery pressures, $SpO_2$ central venous pressure (CVP), peripheral & central temp/ A: 10-s median / S: 2-min median values | Retrospective/ Offline/ Se, Sp, PPV |
| | | Univariate/ Stream-centric/ No SQI/ AD | None |
| Schoenberg [54] | OEM Not specified / 6 patients, total 337 hr | HR, SBP, DBP, $SpO_2$ / Not specified | Real-time/ Real-time/ Se, PPV |
| | | Multivariate/ Stream-centric/ Unknown output score in case of missing data / AD | iMDsoft Clinical Information Systems [101] |
| Zhang [144] | HP Viridia neonatal component monitoring system (Hewlett Packard)/ 11 patients, total 196 hr | HR, PR, RR, SBP, DBP, MBP, $SAO_2$, venous $O_2$ saturation, $O_2$ perfusion/ A: 1 Hz/ S: 1 Hz and derived 1-min averages | Real-time/ Real-time/ Se, Sp, PPV, Acc |
| | | Multivariate/ Stream + patient -centric/ No SQI / AD | None |
| Tsien [145] | SpaceLabs monitor (SpaceLabs Medical, Redmond, WA, USA)/ I: monitored data containing all five signals of interest | HR, RR, $SpO_2$,Invasive SBP, DBP, MBP / A: 1 sample every 5 s | Real-time/ Offline/ AUROC |
| | | Multivariate/ Stream-centric/ No SQI / AD | None |
| Tsien [85] | SpaceLabs monitor (SpaceLabs Medical, Redmond, WA, USA)/ E:cardiac PICU patients/ 35118 minutes | HR, ECG lead number, RR, $SpO_2$, Invasive SBP, DBP, MBP/A & S: 1 sample every 5-6 s | Real-time/ Real-time/ TA, FA |
| | | Not specified/ Stream-centric/ No SQI/ Annotated data for AD | None |
| Monasterio [147] | [1]MIMIC II database / I: $SpO_2$ data / 1616 events annotated on 27 patients | Two ECG leads, Impedance Pneumogram (IP), PPG each at 125 Hz; HR, $SpO_2$ at 1 Hz | Retrospective/ Offline/ Acc, Sp, Se, PPV, NPV |
| | | Multivariate/ Stream-centric/ Specified SQI / AD | None |
| Hoshik [148] | GE Solar 8000M and I and Dash 3000 (GE Healthcare, Milwaukee, USA)/ 1100 patients | Three ECG leads at 240 Hz, Chest Impedance (CI) at 60 Hz, PPG,HR,RR, $SpO_2$ each at 0.5 Hz; alarms, respiratory support, demographics, apnea-bradycardia documentation | Retrospective/ Offline/ FPR, FNR |
| | | Multivariate/ Stream + disease-centric/ Discussed but not specified/ AD + CED | None |
| Nizami [71] | Not mentioned | ECG, PPG | Not evaluated |
| | | Multivariate/Stream + disease-centric/SQI utilized/ AD + CED | Artemis [104] |
| Salatian [26] | OEM Not specified / 1 data segment | BP at 1 Hz | Not discussed/ Offline/ Graphical display |
| | | Univariate/ Stream-centric/ No SQI/ AD | None |

| Author | Theme II: Physiologic Data Source<br>Data Monitor/ Inclusion (I)/ Exclusion (E) criteria/ Sample size | Theme III: Harvested Data<br>Data Type/Acquisition(A)/ Sampling or Storage (S) Frequency<br>Theme IV: Data Analysis<br>Dimensionality/ Focus/ Signal Quality/ Clinical Contribution | Theme V: Clinical Evaluation<br>Data Annotation/ Mode/ Performance Metric<br>Theme VI: Clinical Implementation |
|---|---|---|---|
| Belal [149] | Hewlett Packard Merlin M1064a 1176a/ I: neonates < 2 months old/ 54 patients, total 2426 hr | HR, RR, $SpO_2$/ S: 1 Hz | Partially retrospective & automated markup/ Offline/ Acc, Sp, Se, ROCC |
| | | Multivariate/ Stream + disease-centric/ No SQI/ AD + CED | None |
| Blount [80] | OEM Not specified/ 4 patients | Nursing interventions causing artifacts | Real-time/ Offline/ Not specified |
| | | Not specified/ Stream-centric/ No SQI/ AD | None |
| Quinn [150] | OEM Not specified/ I: 24-29 weeks gestation premature babies in their 1st week of life/ 15 patients x 24 hr | Core and Peripheral temp, DBP, SBP, HR, $SpO_2$, $TpCO_2$, $TpO_2$/ A & S: 1 Hz | Retrospective/ Offline/ AUROC, EER |
| | | Multivariate/ Stream + disease-centric/ Noticed but not discussed/ AD + CED | None |
| Portet [67] | OEM Not specified/ I: preterm infants/ 13 patients x 24 hr | Core and Peripheral temp, DBP, SBP, HR, $SpO_2$, $TpCO_2$, $TpO_2$, humidity of incubator/ A &S: 1 Hz | Retrospective/ Offline / Se, Sp, Acc, k |
| | | Univariate/ Stream + disease-centric/ No SQI/ AD + CED | BabyTalk project [111] |
| Walls-Esquival [31] | OEM Not specified/ I: preterm infants < 30 wks gestation | EEG, ECG, EMG, RR | Real-time with video/ Real-time/ Not applicable |
| | | Multivariate/ Stream-centric/ No SQI / AD | Not applicable |
| Moorman [84] | Not specified | ECG at S: 4 kHz | Real-time / Real-time/ AUROC, Positive Predictive Accuracy |
| | | Multivariate (d)/ Stream + disease-centric/ No SQI/ AD + CED + PD | $HeRO^{TM}$ system [114] |
| Tsien [151] | OEM Not specified/ 274 hr | HR, IBP, $TpCO_2$, $TpO_2$/ A & S: 1 Hz and 1min averages | Retrospective/ Offline/ AUROC |
| | | Multivariate/ Stream-centric/ No SQI/ AD | None |
| Tsien [152] | OEM Not specified/ E: less than four monitored signals / 200 hr | HR, BP, $TpCO_2$, $TpO_2$/ S: 1 sample per min | Retrospective/ Offline/ AUROC |
| | | Multivariate/ Stream-centric/ No SQI/ AD | None |
| McIntosh [64] | Hewlett Packard 78344A (South Queensferry, U.K.)/ I: pneumothorax/ E: infants with birth asphyxia, persistent pulmonary hypertension, requiring inotropic support/ 42 patients | $TpCO_2$, $TpO_2$/ A: 1 Hz  & S: 1-min average | Retrospective/ Offline/ AUROC,PPV, NPV |
| | | Multivariate/ Stream + disease –centric/ No SQI /AD + CED | None |
| Cao [65] | Hewlett Packard 78344 multichannel neonatal monitors/ I: preterm infants/ 10 patients x 10 hr, total 6000 values of $PO_2$, $PCO_2$ | $TpCO_2$, $TpO_2$/ A: 1 Hz  & S: 1-min average | Retrospective/ Offline/ Se, Sp |
| | | Multivariate/ Stream-centric/ No SQI / AD | None |

| Author | Theme II: Physiologic Data Source<br>Data Monitor/ Inclusion (I)/ Exclusion (E) criteria/ Sample size | Theme III: Harvested Data<br>Data Type/Acquisition(A)/ Sampling<br>or Storage (S) Frequency<br>Theme IV: Data Analysis<br>Dimensionality/ Focus/ Signal Quality/ Clinical Contribution | Theme V: Clinical Evaluation<br>Data Annotation/ Mode/<br>Performance Metric<br>Theme VI: Clinical Implementation |
|---|---|---|---|
| Karlen [90] | (i) Capnobase database using Nellcor pulse oximeter & S/5 Collect software (Datex-Ohmeda, Finland) (ii) Complex System Laboratory database/ I: general anesthesia/ (i) 124 patients x 120 sec + 42 patients x 480 sec (ii) 2 patients | PPG/ (i) A: 100 Hz & S: 300 Hz | Retrospective/ Offline/ Se, PPV |
|  |  | Univariate/ Stream-centric/ Developed  SQI/ AD | None |
| Koeny [153] | OEM Not specified/ 17 patients | HR, NIBP/ A: 0.33 Hz/ S: 0.0167 Hz | Real-time/ Offline/ Graphical display |
|  |  | Univariate/ Stream-centric/ No SQI/ AD | None |
| Karlen [91] | Capnobase database using S/5 Collect software (Datex-Ohmeda, Finland)/ I: general anesthesia/ 42 patients | ECG, PPG, RR/ A: ECG at 300 Hz, PPG at 100 Hz; RR at 25 Hz; S: 300 Hz | Retrospective/ Offline/ Error, Power, Robustness |
|  |  | Multivariate/ Stream-centric/ Discussed but  not specified/ AD | None |
| Schmid [36] | Kappa XLT monitor (Dräger), Zeus anesthesia workstation (Dräger), Nortis MedLink, (Lubeck, Germany), Erasmus MC eData TapeRec, (Rotterdam, The Netherlands)/ I: anesthesia for elective cardiac surgery (aortocoronary bypass grafting and valve surgery)/ 25 patients | ECG, PPG, SBP, MAP, DBP, CVP, LAP, $SpO_2$, Temp, MV, RR, PAW, $CO_{2\,exp}$, $CO_{2\,insp}$, Isoflurane$_{Insp}$, Isoflurane$_{exp}$, alarms/Not specified | Retrospective with video/ Offline/ Alarm Validity |
|  |  | Univariate/ Stream-centric/ No SQI/ AD | None |
| Yang [53] | GE S/5 Monitor/ 40 patients | $EtCO_2$, MAP, MV exp, NIBP mean/ A & S: 1 sample every 5 sec | Retrospective/  Offline/ TPR, FPR |
|  |  | Univariate/ Stream-centric/ No SQI/ AD | None |
| Yang [58] | OEM Not specified/ 10 patients | NIBP mean/ A & S: 1 sample every 5 sec | Retrospective/  Offline/ TPR, FPR |
|  |  | Univariate/ Stream-centric/ No SQI/ AD | None |
| Ansermino [155] | GE S/5 Monitor/ 47 surgeries x 1 hr (19 children, 28 adults) | HR, NIBP mean, $SpO_2$, $EtCO_2$, MV exp, RR/ Not specified | Real-time/ Offline/ Se, PPV, NPV, TPR, FPR |
|  |  | Univariate/ Stream-centric/ No SQI/ AD | None |
| Yang [154] | GE S/5 Monitor/ E: fibrillation and other instances of abnormal heart rhythm/ 2 patients | HR, PR, IBP/ A: 1 sample every 5 sec | Retrospective/ Simulated data & [155]/ RMSE |
|  |  | Multivariate/ Stream-centric/ No SQI / AD | None |
| Hoare [72] | Local hospital database, OEM Not specified / 245 cases | HR/ A & S: 1 sample every 30 sec | Retrospective/ Offline/ AUROC, FPR, PPV |
|  |  | Univariate/ Stream-centric/ No SQI/ AD | None |
| Gostt [156] | Datex AS/3 Anaesthesia Monitor, Nellcor N-200/ 9 paediatric and 11 adult patients | HR, PR/ A: 1 sample every 5 sec, but not always due to data capture issues | Retrospective/ Real-time/ Acc, Se, Sp |
|  |  | Multivariate analysis for univariate problem/ Stream + disease-centric/ No SQI/ AD + CED | None |

| Author | Theme II: Physiologic Data Source Data Monitor/ Inclusion (I)/ Exclusion (E) criteria/ Sample size | Theme III: Harvested Data Data Type/Acquisition(A)/ Sampling or Storage (S) Frequency Theme IV: Data Analysis Dimensionality/ Focus/ Signal Quality/ Clinical Contribution | Theme V: Clinical Evaluation Data Annotation/ Mode/ Performance Metric Theme VI: Clinical Implementation |
|---|---|---|---|
| Sittig [157] | OEM Not specified / I: during open heart surgery/ 1 patient | HR, HR from ABP, Pulmonary artery mean pressure, MAP/ A: 1 sample every 2 min S: 1 sample repeated every sec | Not specified/ Real-time/ Graphical evaluation |
| | | Univariate/ Stream-centric/ No SQI/ AD | None |
| Navabi [158] | 5 kinds of Datascope monitors, Nellcor PO, Critikon VRP respiratory monitor/ 21 surgical cases | HR, EtCO$_2$, Inspired O$_2$, NIBP, SAO$_2$, HR, Inspired & expired lung volumes, RR, MAP, MV/ A: EtCO$_2$, Inspired O$_2$, NIBP, HR, SAO$_2$: 0.2 Hz, Inspired & expired lung volumes, RR, MAP, MV: 0.0167 Hz | Not specified/ Real-time/ CDR, FARR |
| | | Multivariate/ Stream-centric/ No SQI / AD | None |
| Ebrahim [22] Feldman [23] | SpaceLabs Medical Gateway, SpaceLabs Medical PC2/ 12 OR, 60 adult ICU, 13 PICU patients, each ICU record is 4 hours long | HR, BP, PR, SpO$_2$/ A & S: HR, PR reading every 3- 5 s | Real-time/ Real-time in [23]/ False and missed alarms |
| | | Multivariate analysis for univariate problem/ Stream centric/ No SQI/ AD+PD | None |
| Clifford [34] | [2]Sana Project database / 30,000 x 10 sec | Twelve lead ECG/ A & S: 500 Hz | Retrospective/ Offline/ Acc, ROCC |
| | | Multivariate type (d)/ Stream-centric/ Developed SQI / AD | None |
| Redmond [44] | TeleMedCareHealth Monitor (TeleMedCare Pty Ltd Sydney, Australia)/ I: home dwelling patients with chronic obstructive pulmonary disease and/or congestive heart failure/ 288 patients, total 300 recordings | Single Lead I-ECG/ S: 500 Hz | Retrospective/ Offline/ Se, Sp, Acc, κ |
| | | Univariate/ Stream-centric/ SQI can be developed/ AD | None |
| Reisner [159] | Propaq 206EL monitors (Protocol Systems, Beaverton, Ore)/ I: prehospital trauma adult patients/ 671 patients | ECG, RR, HR, SBP, DBP/ A & S: ECG at 182 Hz; HR, RR at 1 Hz | Retrospective/ Offline/ AUROC |
| | | Multivariate / Stream + disease-centric/ No SQI/ AD + CED | None |
| Martinez-Tabares [160] | (i) INNOVATEC S.L. (Spain), CELBIT LTDA. ELECTRODOCTOR (Colombia), Lionheart 1 (BIO-TEK) simulator, QRS-Card (Pulse Biomedical Inc) (ii) [2]Sana Project database/ (ii) 1000 x 12 sec | Twelve lead ECG/ A & S: 500 Hz | Retrospective/ Offline/ Se, Sp, Acc, ROCC |
| | | Multivariate / Stream-centric/ Developed SQI / AD | None |
| Silva [161] | [1]MIMIC II database/ 1361 epochs | Multi-lead ECG, PPG, ABP, Resp, CVP/ S: 125 Hz | Retrospective/ Offline/ AUROC |
| | | Multivariate/ Stream-centric/ Developed SQI / AD | None |

| Author | Theme II: Physiologic Data Source<br>Data Monitor/ Inclusion (I)/ Exclusion (E) criteria/ Sample size | Theme III: Harvested Data<br>Data Type/Acquisition(A)/ Sampling or Storage (S) Frequency<br>Theme IV: Data Analysis<br>Dimensionality/ Focus/ Signal Quality/ Clinical Contribution | Theme V: Clinical Evaluation<br>Data Annotation/ Mode/ Performance Metric<br>Theme VI: Clinical Implementation |
|---|---|---|---|
| Hayn [61] | [2]Sana Project database / 2000 x 10 sec | Twelve lead ECG/ A & S: 500 Hz | Retrospective/ Offline/ Se, Sp, Acc |
| | | Multivariate type (d)/ Stream-centric/ Developed SQI / AD | None |
| Jekova [50] | [2]Sana Project database / 1500 x 10 sec | Twelve lead ECG/ A & S: 500 Hz | Retrospective/ Offline/ Se, Sp, ROCC |
| | | Multivariate type (d)/ Stream-centric/ Developed SQI / AD | None |
| Di Marco [62] | [2]Sana Project database / 1498 x 10 sec | Twelve lead ECG/ A & S: 500 Hz | Retrospective/ Offline/ Acc, ROCC |
| | | Multivariate type (d)/ Stream-centric/ Developed SQI / AD | None |
| Johannesen [63] | [2]Sana Project database / 1500 x 10 sec | Twelve lead ECG/ A & S: 500 Hz | Retrospective/ Offline/ Acc, ROC curves |
| | | Multivariate type (d)/ Stream-centric/ Developed SQI / AD | None |
| Addison [18] | Nell-1 oximeter module, (Covidien-Nellcor, Boulder, CO, USA) with a Nellcor Max-A disposable probe, Datex-Ohmeda CardioCap/S5/ 139 healthy volunteers x 8 min | PPG, EtCO$_2$/ A: PPG at 75.7 Hz | Real-time/ Real-time/ Root Mean Square Difference (RMSD) |
| | | Univariate/ Stream-centric/ No SQI/ AD + PD | RR$_{OXI}$ (Covidien-Nellcor,Boulder,CO, USA) |
| Lázaro [17] | Biopac OXY100C,ECG100C, RSP100C sensor, TSD201 transducer & Finometer / 17 subjects x 9 min | PPG at 250 Hz, ECG leads I, III & precordials at 1000 Hz, respiratory signal r(n) at 125 Hz, BP at 250 Hz | Not specified/ Offline/ Inter-subject mean & standard deviation |
| | | Multivariate/ Stream-centric/ No SQI / AD + PD | None |
| Otero [100] | Polysomnographic device, ( Nicolet Biomedical Inc.)/ I: sleep study/ 10 patients, total 59 hr 10 min | Respiratory airflow, SpO2/ A: 68.25 Hz; S: 4 Hz | Retrospective/ Offline, TRACE [174]/ CDR, FPR |
| | | Multivariate / Stream + disease-centric/ No SQI/ AD + CED | None |
| Bsoul [162] | PhysioNet Apnea-ECG database/ 35 subjects | Polysomnography and ECG/ A: ECG at 100 Hz, S: ECG at 250 Hz | Retrospective/ Offline/ F-measure, Se, Acc, |
| | | Multivariate / Stream + disease-centric/ No SQI/ AD + CED + PD | None |
| Xia [163] | [2]Sana Project database/ 1000 x 10 sec | Twelve lead ECG/ S: 500 Hz | Retrospective/ Offline/ Acc, Se, Sp |
| | | Multivariate type (d)/ Stream-centric/ Discussed SQI / AD | None |
| Zaunseder [164] | [2]Sana Project database/ 1000 x 10 sec | Twelve lead ECG/ S: 500 Hz | Retrospective/ Offline/ Acc |
| | | Multivariate type (d)/ Stream-centric/ Developed SQI / AD | None |

| Author | Theme II: Physiologic Data Source Data Monitor/ Inclusion (I)/ Exclusion (E) criteria/ Sample size | Theme III: Harvested Data Data Type/Acquisition(A)/ Sampling or Storage (S) Frequency Theme IV: Data Analysis Dimensionality/ Focus/ Signal Quality/ Clinical Contribution | Theme V: Clinical Evaluation Data Annotation/ Mode/ Performance Metric Theme VI: Clinical Implementation |
|---|---|---|---|
| Kužílek [165] | [2]Sana Project database/ Not clear | Twelve lead ECG/ S: 500 Hz | Retrospective/ Offline/ Not clear |
| | | Multivariate type (d)/ Stream-centric/ No SQI / AD | None |
| Sukor [29] | MLT1020FC Reflection mode infrared finger probe, Differential bio-amplifier ST4400, PowerLab data acquisition system (ADInstruments, Sydney, Australia)/ 13 healthy subjects | PPG, ECG/ A: 1 kHz | Retrospective & Real-time/ Offline/ Se, Sp, Acc, к |
| | | Multivariate validation of a univariate classifier/ Stream-centric/ SQI can be developed/AD + PD | None |
| Acharya [166] | Sleep staging signals Grass amplifiers (Astro-Med Inc., USA), other OEM not specified/ 25 subjects with suspected disease, 14 normal subjects | ECG/ A & S: 256 Hz | Real-time PSG annotations by a clinician/ Offline/ Acc, Se, Sp, PPV using confusion matrix |
| | | Univariate/ Stream + disease-centric/ No SQI/ AD + CED | None |
| Khandoker [167] | OEM Not specified/ I: sleep apnoea & E: cardiac history/ 29 patients | EEG, electrooculograms, ECG,leg movements, body positions, thoracic & abdominal wall expansion, oronasal airflow, SpO₂/ S: 250 Hz for ECG, S: 32 Hz for thoracic and abdominal wall expansion | Polysomnography annotations/ Offline/ Acc, Se, Sp |
| | | Multivariate validation of a univariate classifier/ Stream + disease-centric/ No SQI/ AD + CED | None |
| Nizami [168] | PhysioNet databases: nsrdb; MIT-BIH: nsr2db, mitdb, svdb; BIDMC chfdb, chf2db/ I: various records with normal sinus rhythm, arrhythmia or congestive heart failure/100 patient records | RR Interval / Not applicable | Retrospective/ Offline/ Acc |
| | | Univariate/ Stream + disease-centric/ No SQI/ AD + CED | None |
| Nemati [169] | Peripheral Arterial Tonometry (PAT) measurement (Itamar Medical, Isreal), other OEMs not specified/ I: apnoea-hypopnoea index (AHI) of 0 – 69.3 events/hour 30 patients x 6-8 hr | ECG, 4 channels of respiratory rate (from chest & abdomen plethysmograph, nasal & oral thermistor, nasal pressure), PAT / A & S: 500 Hz, except PAT at 100 Hz | Polysomnography annotations/ Offline/ Signal-to-Noise Ratio |
| | | Multivariate/ Stream-centric/ Developed and implemented SQI/ AD | None |
| Alvarez [170] | Polysomnograph (Alice 5, Respironics, Philips Healthcare, The Netherlands), Nonin PureSAT PO (Nonin Medical, Plymouth, MN, USA)/ I: daytime hypersomnolence, loudsnoring, nocturnal choking and awakening, apneic events/ E: other sleep disorders e.g., insomnia, parasomnia, narcolepsy/ 148 patients x 8 hr | SaO₂/ S: 1 Hz | Real-time Polysomnography annotations by a clinician/ Offline/ Acc, Se, Sp, ROCC |
| | | Univariate/ Stream + disease-centric/ No SQI/ AD + CED | None |

| Author | Theme II: Physiologic Data Source Data Monitor/ Inclusion (I)/ Exclusion (E) criteria/ Sample size | Theme III: Harvested Data Data Type/Acquisition(A)/ Sampling or Storage (S) Frequency / Theme IV: Data Analysis Dimensionality/ Focus/ Signal Quality/ Clinical Contribution | Theme V: Clinical Evaluation Data Annotation/ Mode/ Performance Metric / Theme VI: Clinical Implementation |
|---|---|---|---|
| Chen [69] | Propaq EncoreR 206EL monitor (ProtocolR Systems Inc.)/ I: hemorrhaged patients receiving blood with documented injury/ E: insufficient documentation of injury / 823 patients | at least one of HR, RR, DBP, SBP, $SaO_2$ A & S: 1 Hz, except BP taken intermittently | Retrospective/ Offline/ AUROC |
| | | Multivariate / Stream + disease-centric/ Developed SQI / AD + CED | None |
| Kostic [83] | ECG recorders: Signalife Fidelity 100, NorthEast DR180+, Mortara ELI 200, HP Page Writer 1700A & GE MAC 5000/ I: percutaneous coronary intervention (PCI)/ 54 patients | ECG/ S: 720 Hz | Real-time/ Real-time/ Qualitative, graphical assessment |
| | | Univariate/ Stream + disease-centric/ No SQI/ AD + CED | None |
| Yu [173] | Propaq EncoreR 206EL monitor (ProtocolR Systems Inc.)/ I: trauma patients during helicopter transport to a hospital/ 726 patients x 25 mins | ECG waveform at 182 Hz, PPG waveform at 91 Hz, HR, PR at 1 Hz | Retrospective/ Offline/ AUROC |
| | | Multivariate / Stream + disease-centric/ Developed SQI / AD + CED | None |
| Gil [171, 172] | (i) ECG Apnea Data Base from Physionet; (ii) Digital polygraph EGP800 (Bitmed), COSMO $EtCO_2/SpO_2$ Monitor(Novametrix, Medical Systems)/ (i) I: children suspected of having Obstructive sleep apnea syndrome (OSAS)/ (ii) E: lack of manual annotations, unacceptable respiratory flow signal quality, doubtful clinical diagnosis/ (i) 70 adults (ii) 26 children | Polysomnography data: a chin electromyogram, 6 EEG & 2 electro-oculogram channels, ECG, air flow, respiratory plethysmography, PPG, $SaO_2$, $CO_2$/ S: all signals at 100 Hz | Retrospective/ Offline/ Se, Sp, PPV |
| | | Multivariate / Stream + disease-centric/ No SQI / AD + CED | None |
| Tarassenko [66] | OEM not specified / I: patients monitored for at least 24 hr after a myocardial infarct and again for a few hours 5 days later; patients with severe heart failure; patients with acute respiratory problems; elderly patients with hip fracture/ 150 patients x 24 hr | HR, RR,$SpO_2$,temp, BP/ A & S: 1 sample every min | Retrospective & Real-time/ Real time, BioSign system/ Percentage of True alarms |
| | | Multivariate/Stream + patient-centric/ No SQI/ AD | None |

[1]PhysioNet MIMIC II database recorded using Philips Intellivue MP-70 Medical Systems; [2]PhysioNet database: Sana Project, OEM not specified; [3]PhysioNet's MIMIC database recorded using Hewlett Packard CMS "Merlin" bedside monitor. *Abbreviations and acronyms*: Accuracy (Acc); Sensitivity (Se); Specificity (Sp); Receiver Operating Characteristic Curve (ROCC); Area Under the ROC Curve (AUROC); Positive Predictive Value (PPV); Negative Predictive Value (NPV); Correct Detection Rate (CDR); False Positive Rate (FPR); False Negative Rate (FNR); True Alarm Rate (TA); False Alarm Rate (FA); Root Mean Square Error (RMSE); False Alarm Reduction Rate(FARR); Equal Error Rate (EER); Cohen's kappa coefficient (k); Electrooculograms (EOG).